# High-temperature intrinsic ferromagnetism in the (In,Fe)Sb semiconductor

A.V. Kudrin[1,*], Yu.A. Danilov[1], V.P. Lesnikov[1], M.V. Dorokhin[1], O.V. Vikhrova[1], D.A. Pavlov[1], Yu.V. Usov[1], I.N. Antonov[1], R.N. Kriukov[1], A.V. Alaferdov[2] and N.A. Sobolev[3,4]

[1]*Lobachevsky State University of Nizhny Novgorod, Gagarin av. 23/3, 603950 Nizhny Novgorod, Russia*
[2]*Center for Semiconductor Components and Nanotechnologies, State University of Campinas, Campinas, 13083-870 SP, Brazil*
[3]*Department of Physics and I3N, University of Aveiro, 3810-193 Aveiro, Portugal*
[4]*National University of Science and Technology "MISiS", 119049 Moscow, Russia*
*kudrin@nifti.unn.ru

(In,Fe)Sb layers with a Fe content up to 13 at.% have been grown on (001) GaAs substrates using the pulsed laser deposition. Transmission electron microscopy shows that the layers are epitaxial and free of second-phase inclusions. The observation of hysteretic magnetoresistance curves at temperatures up to 300 K and the investigations of magnetic circular dichroism reveal that the Curie point lies above room temperature. The resonant character of magnetic circular dichroism confirms the intrinsic ferromagnetism in the (In,Fe)Sb matrix. We suggest that the ferromagnetism of the (In,Fe)Sb matrix is not carrier-mediated and is apparently determined by the mechanism of superexchange interaction between Fe atoms.

## I. INTRODUCTION

Recently, a number of experimental works have shown the appearance of a ferromagnetic ordering in III-V semiconductor hosts heavily doped with Fe atoms, in particular, (In,Fe)As [1,2], (Ga,Fe)Sb [3,4], (Al,Fe)Sb [5]. The MBE (molecular-beam epitaxy) (Ga,Fe)Sb layers demonstrated ferromagnetism in magnetic, magneto-optic and magnetotransport properties up to 340 K [4]. According to the cited experimental works, the sizeable amount of Fe impurity can be introduced into the InAs, GaSb and AlSb semiconductor hosts without the formation of second-phase inclusions (up to 9% relative to the In content for (In,Fe)As, up to 25% relative to the Ga content in (Ga,Fe)Sb and up to 10% relative to the Al content in (Al,Fe)Sb). The origin of the ferromagnetism in these magnetic semiconductors remains insufficiently clear. It is suggested that in the narrow-bandgap (In,Fe)As and (Ga,Fe)Sb semiconductors the ferromagnetism is associated with the mechanism of indirect *s,p-d* exchange interaction as in the case of manganese doping. However, in contrast to manganese, the electrical activity of iron in such materials remains a controversial issue. In the wide-bandgap semiconductor (Al,Fe)Sb the charge-carrier concentration at low temperatures (less than $10^{16}$ cm$^{-3}$) is too low for carrier-mediated exchange interaction [5]. For this case, the mechanism of short-range superexchange between Fe ions has been assumed for the explanation of the intrinsic ferromagnetism [5].

In this work we present the results of investigations of deposited (In,Fe)Sb epitaxial layers manifesting ferromagnetic properties at least up to room temperature (RT).

## II. EXPERIMENTAL

The (In,Fe)Sb layers were grown by pulsed laser sputtering of semiconducting InSb and metallic Fe targets in a vacuum chamber with a background gas pressure of about $2\times10^{-5}$ Pa. The presence of an additional Sb target allowed us to introduce an additional amount of antimony during the sputtering process. A semi-insulating (001) GaAs was used as a substrate. The growth temperature ($T_g$) was varied in the range of 150 – 250ºC. The Fe content was characterized by the technological parameter $Y_{Fe} = t_{Fe}/(t_{Fe}+t_{InSb})$, where $t_{Fe}$ and $t_{InSb}$ are the ablation times of the Fe and InSb targets, respectively. Structural properties were investigated by high-resolution cross-sectional transmission electron microscopy (TEM). The surface morphology of the layers was investigated by both atomic force microscopy (AFM) and scanning electron microscopy (SEM). The distribution of constituent elements was obtained by energy-dispersive X-ray spectroscopy (EDS) during TEM and SEM investigations. The dc magnetotransport measurements were carried out in a van der Pauw geometry from 15 to 300 K in a closed-cycle He cryostat. The Seebeck coefficient α = –$\Delta V/\Delta T$ was measured in the range from 300 – 330 K. Reflectivity spectra were obtained at room temperature in the spectral range of 1.6 – 6 eV. The investigations of the magnetic circular dichroism (MCD) were performed in the cryostat in the spectral range from 1.55 to 2.84 eV using a xenon arc lamp as a light source.

In this study we present results for the following structures: an undoped InSb layer grown at 250ºC (structure 250-0), (In,Fe)Sb layers with a Fe content $Y_{Fe} = 0.17$ grown at 150, 200 and 250ºC (structures 150-17, 200-17, and 250-17), a (In,Fe)Sb layer with $Y_{Fe} = 0.17$ grown at 200ºC upon sputtering of an additional Sb target during the layer growth (structure 200(Sb)-17) and a (In,Fe)Sb layer with $Y_{Fe} = 0.08$ grown at 250ºC (structure 250-8).

## III. RESULTS

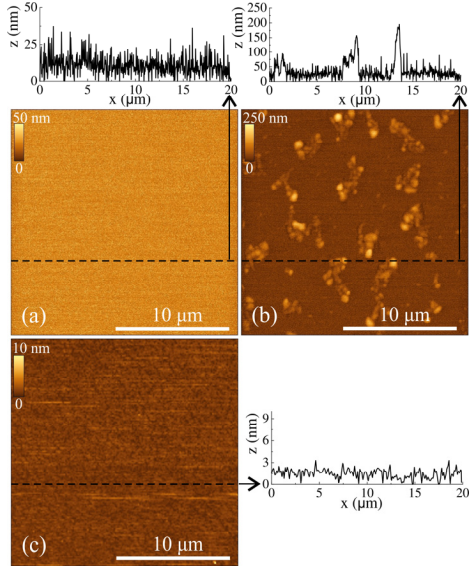

FIG. 1. AFM images and height profiles. (a) the InSb layer (structure 250-0). (b) the (In,Fe)Sb layer (structure 250-17). (c) the (In,Fe)Sb layer grown upon sputtering of an additional Sb target during the layer growth (structure 200(Sb)-17).

Figure 1 shows the AFM surface morphology of structures 250-0 (Fig. 1(a)), 250-17 (Fig. 1(b)) and 200(Sb)-17 (Fig. 1(c)). The undoped InSb layer has a fairly smooth surface with a root mean square (rms) roughness of about 7 nm (Fig. 1(a)). It should be noted that a peculiarity of the used growth method is a uniform coverage of the (001) GaAs substrate by the InSb layer and the absence of a pronounced three-dimensional island growth. On the contrary, for MBE grown InSb layers, a complete coverage of the (001) GaAs substrate is achieved only after deposition of ≈ 300 monolayers (97 nm) [6] as a consequence of the island growth mode at the initial stages of the InSb layer growth due to a large lattice mismatch $\Delta a/a$ between GaAs and InSb (14.6 %). A feature of the morphology of the (In,Fe)Sb layers that were grown without co-sputtering of an additional Sb target (Fig. 1(b)) is the presence on the surface of protuberances having the form of elongated pedestals (with a characteristic lateral size of ~ 1 – 3 μm, a height of ~ 50 – 60 nm) exhibiting peaks (with a base diameter of ~ 500 nm and a height of ~ 180 nm). The volume of the protuberances risen above the (In,Fe)Sb surface amounts to ~ 13% relative to the volume of the (In,Fe)Sb layer (taking into account the thickness of the (In,Fe)Sb layer of ≈ 40 nm, see the TEM investigations below). This value is comparable with the technological content of the introduced Fe ($Y_{Fe}$ = 0.17 for structure 250-17). The protuberances on the (In,Fe)Sb surface are areas of indium surface segregation. The appearance of the indium islands on the surface is a consequence of the replacement of In atoms by Fe ones in the (In,Fe)Sb matrix, which under conditions of Sb deficiency leads to the displacement of In on the surface. A similar appearance of In droplets on the InSb surface was observed in InSb/GaAs structures grown by MBE with an excess flux of indium. [7]. Figure 2 presents a SEM image of a protuberance on the surface of structure 200-17 and the EDS mapping of In, Sb, and Fe. The EDS investigations confirm that the protuberances are areas of the In surface segregation. Also in the protuberances, a certain Sb concentration increase is observed. At the same time, the Fe concentration enrichment within the protuberance was not revealed. The emergence of the protuberances can be suppressed by the introduction of an additional amount of Sb in the growing (In,Fe)Sb layer. Figure 1(c) shows the AFM surface morphology of structure 200(Sb)-17 that was grown upon sputtering of an additional Sb target during the growth process. The surface of structure 200(Sb)-17 is very smooth with a rms roughness of about 0.5 nm.

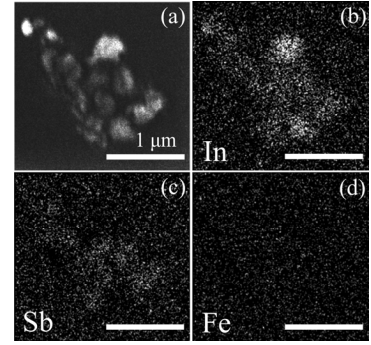

FIG. 2. (a) SEM image of a protuberance on the surface of structure 200-17. (b) – (c) EDS mapping of In, Sb, and Fe.

Figure 3(a) shows a cross-section TEM image of structure 250-17. The overview TEM image was obtained from a 400 nm long region located between the In protuberances discussed above. The image demonstrates a quite smooth (In,Fe)Sb layer with a thickness of about 40 nm without evident second-phase inclusions. Figure 3(b) shows a high-resolution TEM (HRTEM) image of a 170 nm long region. The image also does not reveal any second-phase inclusions. The HRTEM image allows us to obtain a fast Fourier transform (FFT) diffraction pattern (the right part of Fig. 3(b)) of the whole HRTEM image presented in Fig. 3(b). The FFT diffraction pattern contains diffraction spots corresponding to zinc-blende type lattices of the (In,Fe)Sb layer (the inner spots) and the GaAs substrate (the outer spots) and does not contain additional spots from any other crystalline phase. The pattern reveals an epitaxial orientation relationship between the (In,Fe)Sb layer and the GaAs substrate. Figure 3(c) exhibits the HRTEM image at a higher magnification. Due to the large lattice mismatch between the (In,Fe)Sb and GaAs matrixes, a large number of stacking faults are present on the {111} planes. The stacking faults appear in the image as the assemblage of straight lines at an angle of ≈ 70º with respect to each other. The appearance of such stacking faults is typical of epitaxial III-V layers grown on substrates with a large lattice mismatch, in particular, of GaAs layers on Si [8] and InSb layers on GaAs [9]. The HRTEM image and the FFT diffraction pattern of the image for the (In,Fe)Sb layer (the right-hand part of Fig. 3(c)) do not reveal the presence of any visible second-phase inclusions with lattice parameters different from that of the zinc-blende (In,Fe)Sb matrix. The lattice mismatch between the (In,Fe)Sb layer and the GaAs substrate in the growth direction detected from HRTEM images equals ≈ 14.9%, which coincides well with the $\Delta a/a$ value for the InSb and GaAs monocrystals. At the same time, the $\Delta a/a$ value in the plane of the (In,Fe)Sb layer is equals ≈ 10.9%, consequently, the (In,Fe)Sb layer is compressively strained. Figure 3(d) shows the EDS mapping of In, Sb, and Fe in sample 250-17. The data reveal a rather uniform distribution of the elements within the (In,Fe)Sb layer. The Fe content detected by EDS is equal to about 13±1 at.%. This value is smaller than the technological $Y_{Fe} = 0.17$ apparently due to the lower ablation speed for the metallic Fe target.

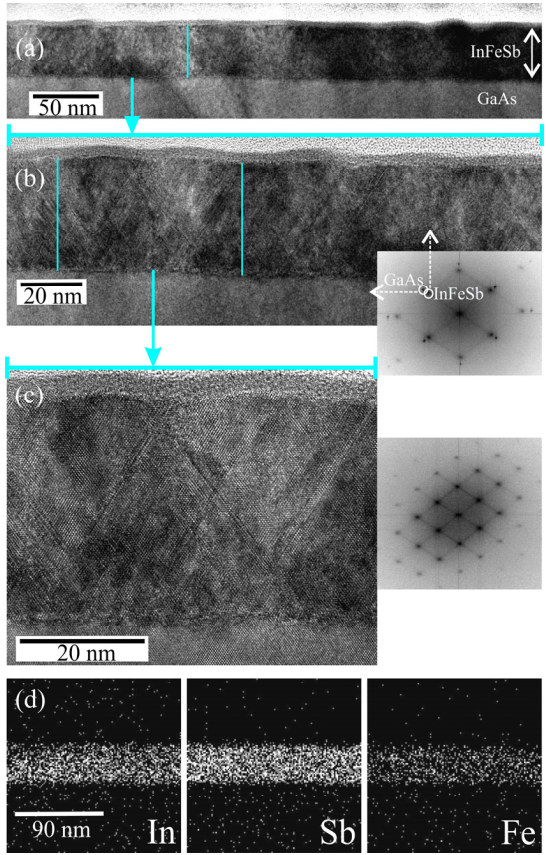

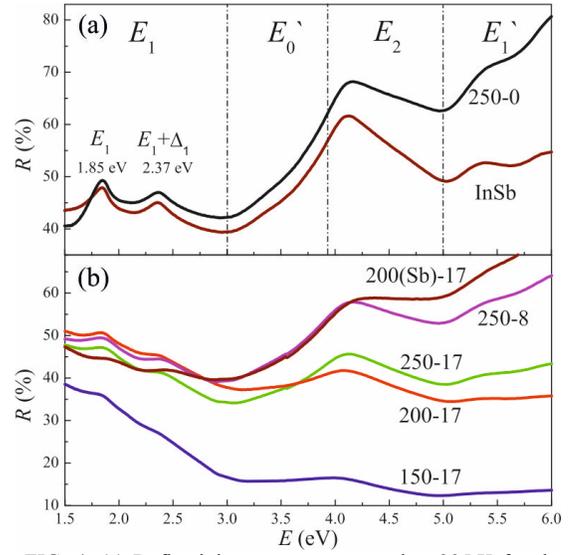

FIG. 3. Results of cross-section TEM investigations of the (In,Fe)Sb/GaAs structure 250-17. (a) Overview TEM image of a 400 nm long region. (b) High resolution TEM image of a 170 nm long region (at the right, a FFT diffraction pattern of the whole image). (c) High resolution TEM image at a higher magnification (at the right, a FFT diffraction pattern of the (In,Fe)Sb part of the image). (d) EDS mapping of In, Sb and Fe in the structure.

FIG. 4. (a) Reflectivity spectra measured at 295 K for the undoped InSb layer and the bulk InSb. (b) Reflectivity spectra taken at 295 K for the (In,Fe)Sb layers.

The results of the mapping and the HRTEM investigations allows us to conclude that the (In,Fe)Sb layers are epitaxial with a uniform distribution of their components.

Figure 4 shows reflectivity spectra measured at 295 K for our structures. Figure 4(a) exhibits the spectra for the undoped InSb layer (structure 250-0) and for the bulk InSb. The reflection spectrum for structure 250-0 coincides with the spectrum for the single crystal InSb and contains features associated with characteristic interband transitions within the Brillouin zone [10]. A doublet in the $E_1$ region (at 1.85 and 2.37 eV) and an intense peak in the $E_2$ region (at 4.13 eV) are the most pronounced features. For the (In,Fe)Sb layers (Figure 4(b)) these characteristic peaks are also well resolved. This confirms the high crystalline quality of the (In,Fe)Sb layers and indicates the conservation of the InSb band structure.

Figure 5 exhibits the temperature dependences of the resistivity $\rho(T)$ for the investigated structures. For all structures the $\rho(T)$ dependences are semiconductor-like – the resistivity increases with decreasing temperature. The investigations of the Hall effect at different temperatures for the undoped InSb layer (structure 250-0 with $n$-type conductivity) showed that the resistivity increase with decreasing temperature is related to both the carrier concentration decrease (from $3\times10^{17}$ cm$^{-3}$ at 295 K to $1.8\times10^{17}$ cm$^{-3}$ at 15 K) and the mobility decrease (from 1240 cm$^2$/V·s at 295 K to 100 cm$^2$/V·s at 15 K). A similar weak temperature dependence of the carrier concentration was noted for thin (30 – 200 nm) epitaxial InSb layers on a (001) GaAs substrate, and it was associated with the presence of a high density of electrically active donor defects [11]. For all (In,Fe)Sb layers the $\rho(T)$ dependences are semiconductor-like (Figure 5). The lowest resistivity in the temperature range from 25 – 295 K is demonstrated by structures 150-17 and 200-17. For structure 200(Sb)-17, the resistivity is 5 – 10 times higher (at different temperatures) than that of structure 200-17. Structure 250-8 has the highest resistivity. For the (In,Fe)Sb layers, it is not possible to trace the temperature evolution of the carrier concentration and mobility separately since the anomalous Hall effect (AHE) dominates in the (In,Fe)Sb layers.

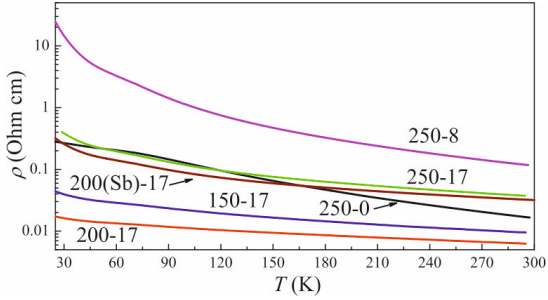

FIG. 5. Temperature dependences of the resistivity for the investigated structures.

Figure 6(a-e) shows the Hall resistance dependences on an external magnetic field ($R_H(B)$) at 300 an 77 K for structures 150-17, 200-17, 250-17, 200(Sb)-17 and 250-8.

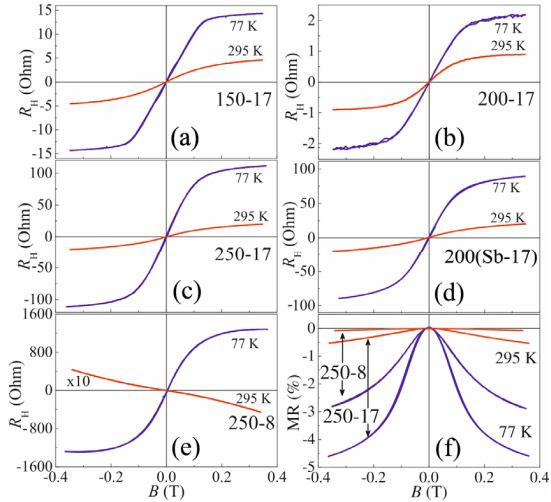

FIG. 6. (a-e) $R_H(B)$ dependences taken at 77 and 295 K for structures 150-17, 200-17, 250-17, 200(Sb)-17 and 250-8. (f) Magnetoresistance curves taken at 77 and 295 K for structures 250-8 and 250-17 ($B$ is perpendicular to the sample plane).

The $R_H(B)$ dependences reveal a clear manifestation of the anomalous Hall effect up to RT – the dependences are nonlinear with a saturation in the field of ≈ 0.15 T. An AHE feature in the investigated (In,Fe)Sb layers is the AHE sign corresponding to the *p*-type conductivity (Figure 6(a-e)). The manifestation of the AHE does not allow us to make an unambiguous conclusion about the type of conductivity of the material. It is known that in magnetic semiconductors both the anomalous and ordinary Hall effect (OHE) of different signs can be observed simultaneously, in particular, in the (In,Fe)As layers [2,12]. At present, the character of electrical activity of the Fe impurity in InSb is disputable. Data about both the donor and the acceptor character of the Fe impurity in the InSb matrix are present in the literature [13]. A feature of structure 250-8 is the presence of both the anomalous and the ordinary Hall effect with different signs at RT (Figure 6(e)). The OHE corresponding to *n*-type conductivity dominates at 295 K at magnetic fields higher than 0.2 T, which allows us to determine the electron concentration of $1\times10^{18}$ cm$^{-3}$, a value being much higher than the electron concentration in the undoped InSb layer at 295 K ($3\times10^{17}$ cm$^{-3}$). It can be assumed that the introduction of Fe into the investigated InSb layers leads to the appearance of additional electrically active donor defects. For structure 250-8 the electron concentration at 295 K ($1\times10^{18}$ cm$^{-3}$) is much higher than that in the intrinsic InSb at room temperature (~ $2\times10^{16}$ cm$^{-3}$). This finding allows us to suggest that the transition to the predominance of the *n*-type ordinary Hall effect at RT for structure 250-8 (Figure 6(e)) is not due to the transition to the intrinsic conductivity of the InSb host, but due to the significant slackening of the AHE, at a temperature (295 K) that is appreciably remote from the Curie point ($T_C$). This is in agreement with the assumption that the obtained (In,Fe)Sb layers are *n*-type and the observed AHE has a sign opposite to that of the OHE. The higher resistivity of structure 250-8 in comparison with structure 250-0 (Figure 5) is apparently associated with the lower electron mobility in structure 250-8, as a result of a stronger electron scattering by donor defects. The resistivity of the structures with $Y_{Fe}$ = 0.17 (150-17, 200-17, 250-17 and 200(Sb)-17) is lower in comparison with structure 250-8 (Figure 5) presumably due to the higher electron concentration associated with a large number of donor defects, caused by the large amount of introduced Fe. The difference in the resistivity of layers 200-17 and 200(Sb)-17 is due to the difference of the density of electrically active defects. For structure 200(Sb)-17, the interaction of the excess indium (segregated on the surface in the case of structure 200-17) with the additional amount of antimony occurs during the growth process. This leads to an increased thickness of the resulting (In,Fe)Sb layer by about 25%. For the equal sputtering time of the Fe target, the Fe concentration in structure 200(Sb)-17 is smaller than that in structure 200-17 (approximately 25% decrease). This leads to a smaller density of electrically active defects in structure 200(Sb)-17 and, consequently, to its increase of resistivity (the resistivity of structure 200(Sb)-17 was calculated taking into account the thickness of the (In,Fe)Sb layer = 50 nm).

The assumption of the *n*-type conductivity at RT in our (In,Fe)Sb layers is confirmed by Seebeck effect measurements. For the undoped InSb layer and for the (In,Fe)Sb layers, the Seebeck coefficient at RT corresponds to the *n*-type conductivity and it is equal to about 10 µV/K. This confirms the assumption that the main charge carriers in the (In,Fe)Sb layers are electrons, although the AHE sign corresponds to the p-type conductivity.

Figure 6(f) shows magnetoresistance (MR = $(\rho(B)-\rho(0))/\rho(0)$) curves for structures 250-8 and 250-17 at 77 and 295 K with the external magnetic field applied perpendicular to the (In,Fe)Sb layer plane. The negative magnetoresistance with a tendency to saturate in magnetic fields > 0.2 T was observed. For all investigated structures the MR curves are similar. The simultaneous manifestation of the anomalous Hall effect and the negative magnetoresistance is an evidence that the charge carrier transport in the (In,Fe)Sb layers is spin-dependent up to RT. The absence of the hysteresis in the $R_H(B)$ and MR curves (in the case of B oriented perpendicular to the plane) is a consequence of the in-plane orientation of the easy magnetization axis due to both the low thickness of the (In,Fe)Sb layers and the presence of compressive strains. In particular, in the case of compressively strained GaMnAs layers (GaMnAs on GaAs substrate) the magnetization vector lies in the layer plane [14,15].

Figure 7 exhibits the MR curves for the (In,Fe)Sb layers in the temperature range from 40 to 295 K for the magnetic field applied in the layer plane. In this case, the MR curves for the layers with $Y_{Fe}$ = 0.17 are hysteretic. This is a consequence of the orientation of the easy magnetization axis and the external magnetic field in the same plane. Figure 8 shows the in-plane MR curves at selected temperatures for B in the range ± 0.04 T.

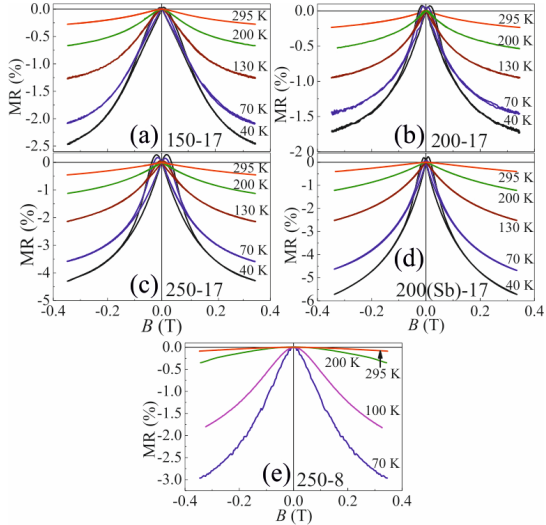

FIG. 7. Magnetoresistance curves measured at various temperatures for structures 150-17, 200-17, 250-17, 200(Sb)-17 and 250-8 for B lying in the sample plane.

For structures 150-17 (Figure 8(a)), 200-17 (Figure 8(b)) and 250-17 (Figure 8(c)) the hysteretic character of the in-plane MR curves remains up to 300 K. The observation of the hysteresis in the magnetoresistance up to 300 K unequivocally reveals that the (In,Fe)Sb/GaAs structures 150-17, 200-17 and 250-17 are ferromagnetic up to RT. For structure 200(Sb)-17 the MR curves are hysteretic up to 170 K (Figure 8(d)). The lower Curie temperature (≈ 170 K) for structure 200(Sb)-17 in comparison with the other structures with $Y_{Fe}$ = 0.17 is associated with a ≈ 25% lower Fe concentration (as discussed above). For structure 250-8 (Figure 7(e)) the hysteresis in the in-plane MR curves at temperatures higher than 70 K was not observed (at temperatures lower than 70 K the MR curves were not obtained due to a large resistivity). This indicates a much weaker ferromagnetic properties of the (In,Fe)Sb layers with a smaller amount of Fe.

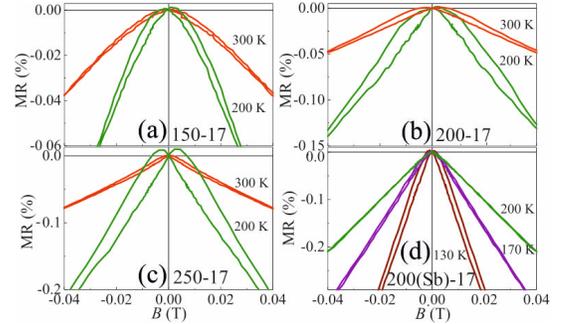

FIG. 8. Magnetoresistance curves measured at various temperatures for structures 150-17, 200-17, 250-17 and 200(Sb)-17 for B varying in the range ± 0.04 T (B lying in the sample plane).

The investigations of the magnetic circular dichroism for structures 200(Sb)-17, 200-17 and 250-0 were performed for the magnetic field applied perpendicular to the sample plane. The light from a xenon arc lamp, after passing through a grating monochromator, was circularly polarized by a combination of a linear polarizer and a quarter-wave plate. The MCD value was defined as $((I^+ - I^-)/(I^+ + I^-)) \times 100\%$, where $I^+$ and $I^-$ are the intensities of circularly polarized light with the right and left circular polarization reflected from the sample's surface. Figure 9 shows the dependences of the $MCD_B$ value (defined as $(MCD(B = +0.3\ T) - MCD(B = -0.3\ T))/2$) on the photon energy of the circularly polarized light (E) for structures 200(Sb)-17, 200-17 and 250-0 at 40 K. For structure 250-0 the $MCD_B$ value is smaller than the MCD measurement error (~ 0.01 %). For structures 200(Sb)-17 and 200-17 the $MCD_B(B)$ dependences are strongly enhanced in the photon energy range from 1.55 to 2.6 eV (Figure 9). For structure 200(Sb)-17 the $MCD_B(B)$ curve reveals two pronounced peaks at ≈ 1.76 and ≈ 2.22 eV.

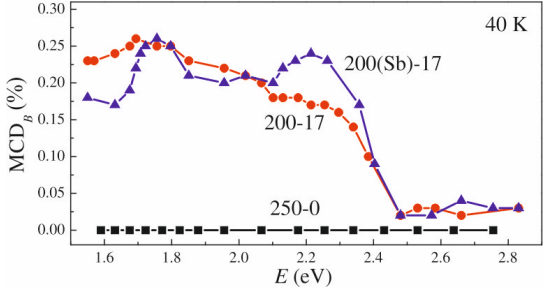

FIG. 9. $MCD_B(E)$ dependences measured at 40 K for structures 200(Sb)-17, 200-17 and 250-0.

These peaks are a consequence of a MCD enhancement for photon energies close to the optical critical points $E_1$ and $E_1+\Delta_1$ of the InSb band structure (1.98 and 2.48 eV for the bulk InSb at 5 K [10], see also Figure 4). For structure 200-17 (with a higher Fe concentration than that in structure 200-17) the peaks are broadened and overlapped (which is typical of heavily doped magnetic semiconductors, in particular for Fe doped ones [3]). For photon energies remote from the optical critical point $E_1+\Delta_1$ (> 2.5 eV) the MCD value is much lower (Figure 9). Thus, the observation of the clear enhancement of the MCD effect at photon energies close to the optical critical points $E_1$ and $E_1+\Delta_1$ confirms the intrinsic ferromagnetism in the obtained semiconductor (In,Fe)Sb layers.

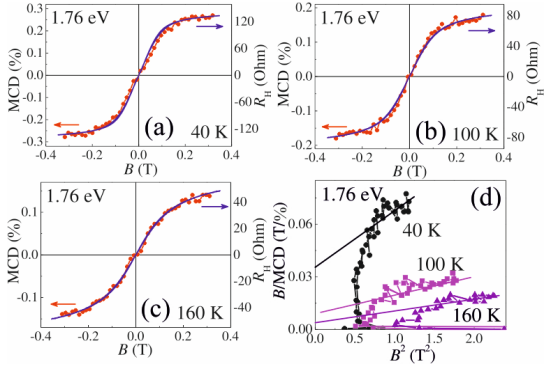

FIG. 10. (a-c) MCD(B) dependences for structures 200(Sb)-17 measured at different temperatures at a photon energy of 1.76 eV. (d) Arrot plots of the MCD(B) curves.

Figure 10 shows MCD dependences in an external magnetic field (MCD(B)) at different temperatures for structure 200(Sb)-17 (Figure 10(a-c)) and corresponding Arrot plots (Figure 10(d)). The MCD(B) curves were obtained using the light with the photon energy 1.76 eV. The MCD(B) dependences are nonlinear with a saturation in a magnetic field of about 0.2 T. The shape of the MCD(B) dependences coincide with the Hall resistance dependences on an external magnetic field at corresponding temperatures for the same structure 200(Sb)-17 (the continuous curves in Figure 10(a-c)). At 295 K the MCD(B) dependence is weak (MCD(Bmax) ~ 0.03%) because the Curie temperature for structure 200(Sb)-17 is below RT. The Arrott plots of the MCD(B) curves reveal $T_C \approx$ 160 K for structure 200(Sb)-17 (this is in accordance with MR investigations (Figure 8)). For structure 200-17, nonlinear MCD(B) dependences are observed up to at least 295 K and also coincide with $R_H(B)$ curves (Figure 11(a-c)). The Arrott plots (Figure 11(c)) as MR investigations (Figure 8) reveal for structure 200-17 a Curie temperature above RT.

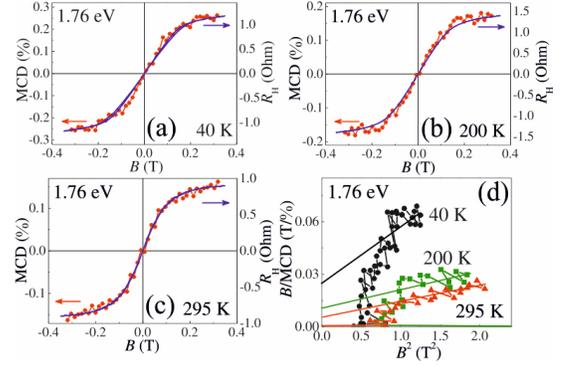

FIG. 11. (a-c) The MCD(B) dependences for structures 200-17 at different temperatures for the photon energy 1.76 eV. (d) The Arrot plots of the MCD(B) curves.

It should be noted that since the structure with a smooth surface (200(Sb)-17, Fig. 1(c)) and the structure with indium islands (200-17, Fig. 2) show qualitatively similar magnetotransport and MCD properties, it can be concluded that the surface protuberances do not affect on the ferromagnetic properties of the (In,Fe)Sb matrix.

### IV. DISCUSSION

The appearance of the ferromagnetism in III-V semiconductors heavily doped with a magnetic impurity is traditionally associated with the mechanism of carrier-mediated exchange interaction between magnetic atoms. The same mechanism of the ferromagnetism was also assumed for semiconductors doped with Fe, in particular, for (In,Fe)As layers [1] with $T_C$ = 70 K (with a carrier (electron) concentration at RT of ~ $2\times10^{19}$ cm$^{-3}$) and (Ga,Fe)Sb layers [4] with $T_C$ = 340 K (with a carrier (hole) concentration at RT of ~ $1\times10^{20}$ cm$^{-3}$). For the explanation of the ferromagnetism in these materials, the assumption, that (according to the vacuum pinning rule) the $d$ level of transition metal atoms is weakly dependent on semiconductor hosts, was used [16,17]. If so, the $d$ level of Fe should lie in the bandgap near the bottom of the conduction band for InAs and near the top of the valence band (within the valence band) for GaSb [3]. It is assumed that for the InAs and GaSb hosts with a large concentration of electrons and holes in the conduction and valence band, respectively, such a location of the Fe level contributes to a large $s,p$-$d$ exchange interaction [3]. With the assumption that the vacuum pinning rule is valid for the InSb host and taking into account the GaSb-InSb bands lineup [18], the $d$ level of Fe for InSb should also lie within the valence band. However, in the epitaxial InSb layers the electrically active donor defects occur ([11], structure 250-0 in this study), in contrast to acceptor defects in GaSb. Consequently, the location of the *d* level of Fe in a completely filled valence band should not give an advantage in the occurrence of the *s*,*p-d* exchange interaction, since charge carriers in (In,Fe)Sb are electrons in the conduction band, and their concentration is provided by the donor levels associated with the defects. In this connection, a different mechanism for the origin of ferromagnetism in our (In,Fe)Sb layers, which is not directly related to the carrier-mediated exchange interaction between Fe atoms, can be assumed. For the explanation of the ferromagnetism in the high-resistivity layers of wide-gap semiconductor (Al,Fe)Sb (with a Fe concentration up to 7 at.%), the mechanism of superexchange interaction between Fe atoms located in the second-nearest-neighbor sites of the host was proposed [5]. This mechanism is effective even at a low carrier density. Our (In,Fe)Sb layers, according to EDS, contain up to 13 at. % of Fe, therefore, in our opinion, the mechanism of the superexchange interaction between Fe atoms is quite probable. This mechanism can be supported by the equal Curie temperature for structures 200-17 and 250-17 (estimated from the observation of hysteresis in the MR curves, Figures 8(b) and 8(c)), although in the temperature range from 77 – 300 K the resistivity of structure 250-17 is about an order of magnitude higher than that of structure 200-17. With the assumption that the carrier mobility in structure 200-17 is not higher than in structure 250-17, the much higher resistance of structure 250-17 should be related to a much lower carrier concentration. The observation of a similar value of $T_C$ for these structures (~ 300 K) points to a mechanism different from the carrier-mediated ferromagnetism.

Earlier we have demonstrated that the presence of second-phase ferromagnetic inclusions in a semiconductor matrix (in particular in (In,Mn)As layers) can result in the observation of nonlinear $R_H(B)$ dependences related not to the AHE, but to the Lorentz force appearing in the layer with an inhomogeneous distribution of the current density and with local magnetic fields produced by the inclusions [19]. Inasmuch as the second-phase inclusions are not observed in the obtained (In,Fe)Sb layers, we can conclude that the AHE and the negative magnetoresistance testify the intrinsic ferromagnetism in the semiconductor matrix and the presence of a spin polarization of charge carriers.

## V. CONCLUSION

In summary, *n*-type (In,Fe)Sb layers with a Fe content of up to 13 at.% have been obtained by the pulsed laser deposition in a vacuum. The high-resolution TEM investigations of the structure of the layers revealed that the layers are epitaxial and do not contain any second-phase inclusions. The anomalous Hall effect and the negative magnetoresistance were observed in the layers up to at least room temperature. The observation of the hysteresis in the magnetoresistance curves up to 300 K reveals that the $T_C$ is higher than room temperature. The resonance character of the magnetic circular dichroism confirms the intrinsic ferromagnetism. We assume that the origin of the ferromagnetic properties of the (In,Fe)Sb matrix is the mechanism of superexchange interaction between Fe atoms. Hence, the obtained (In,Fe)Sb is a single-phase zinc-blende crystal with room temperature intrinsic ferromagnetic properties which manifest themselves in the carrier transport.

## ACKNOWLEDGMENTS

This study was supported by the Grant MK-8221.2016.2 and the Ministry of Education and Science of Russian Federation (Projects No. 8.1751.2017/PCh) and by the FCT of Portugal through the Project No. I3N / FSCOSD (Ref. FCT UID / CTM / 50025 / 2013).